\documentclass[
  ]
  {aipproc}

\layoutstyle{6x9}

\begin{document}

\title{Beauty quark and quarkonium production at LHC:
$k_T$-factorization and CASCADE versus data
}


\author{H. Jung}{
  address={DESY, Hamburg, Germany; University of Antwerp, Antwerp, Belgium}}

\author{M. Kr\"amer}{
  address={DESY, Hamburg, Germany}}

\author{A.V. Lipatov}{
  address={SINP, Moscow State University, 119991 Moscow, Russia}
}

\author{\underline{N.P. Zotov}}{
  address={SINP, Moscow State University, 119991 Moscow, Russia}
}

\begin{abstract}
 We present hadron-level predictions  from the Monte-Carlo
generator CASCADE and numerical calculations of beauty quark
and quarkonium production in the  framework of the $k_T$-factorization approach for LHC energies. Our predictions
are compared with the CMS experimental data.

\end{abstract}
\maketitle

%%%%%%%%%%%%%%%%%%%%%%%%%%%%%%%%%%%%%%%%%%%%
%% MAINMATTER
%%%%%%%%%%%%%%%%%%%%%%%%%%%%%%%%%%%%%%%%%%%%

%\section{Introduction}
Our study is motivated by very recent measurement of open beauty quark and $b$-jet production performed by the CMS Collaboration. It was observed that the data tends to be higher than the MC@NLO predictions and that the shape of the pseudo-rapidity distribution is not well described by MC@NLO. The $p_T$-spectra of b-jets are not well discribed too~\cite{CMS, Chio}.

 Recently we have demonstrated reasonable agreement between the $k_T$-factorization predictions~\cite{JKLZ} and the Tevatron data on the $b$-quark, $b\bar b$ di-jets, $B^+$- and $D$-meson
prodiction and also agreement with total set of HERA data for $J/\psi$-mesons~\cite{BLZ}.
Based on these results, here we give the analysis of the CMS data in the framework of the kT-factorization approach.
We produce the relevant numerical calculations in two ways:
\begin{itemize}
\item We will perform analytical parton-level calculations (which are labeled as LZ).
\item The measured cross sections of heavy quark production will be compared also with the predictions of full hadron level Monte Carlo event generator CASCADE~\cite{Jung}.
\end{itemize}

%\section{Theoretical framework}
The basic dynamical quantity of the
$k_T$-factorization approach is the unintegrated
${\mathbf k}_T$-dependent) gluon distribution (UGD)
${\cal A}(x,{\mathbf k}_T^2,\mu^2)$  obtained from the
analytical or numerical solution of the BFKL or CCFM evolution equations. 

The cross section of any physical process is
calculated as a convolution of the partonic cross section $\hat{\sigma}$ and 
the UGD ${\cal A}_g(x,k_{T}^2,\mu^2)$,
which depend on both the longitudinal momentum fraction 
$x$ and transverse momentum $k_{T}$:
\begin{eqnarray}
  \sigma_{pp} =
  \int {\cal A}_g(x_1,k_{1T}^2,\mu^2)\,{\cal A}_g(x_2,k_{2T}^2,\mu^2)
  \hat{\sigma}_{gg}(x_1, x_2, k_{1T}^2, k_{2T}^2,...)\nonumber
  \,dx_1\,dx_2\,dk_{1T}^2\,dk_{2T}^2.
\end{eqnarray}
The partonic cross section $\hat\sigma$ has
to be taken off mass shell (${\mathbf k}_T$-dependent).\\
 It also assumes a modification of their {polarization density matrix.
 It has to be taken in so called BFKL form:
$$
  \sum \epsilon^{\mu} \epsilon^{*\,\nu} = {k_T^{\mu} k_T^{\nu} \over{\mathbf k}_
T^2 }.
$$
Concerning the UPD in a proton, we used two different sets.
First of them is the KMR one~\cite{KMR}.
The KMR approach represents an approximate treatment  of the parton evolution
mainly based on the DGLAP equation and incorpotating  the BFKL effects at the
last step of the parton ladder only, in the form of the properly defined
Sudakov formfactors $T_q({\mathbf k}_T^2,\mu^2)$ and
$T_g({\mathbf k}_T^2,\mu^2)$,
including logarithmic loop coorections.

Second UGD is the CCFM one. The CCFM evolution equation have been solved numerically using Monte-Carlo} method~\cite{JS}.
 In this case UGD are determined  by a convolution
of the non-perturbative starting distribution
${\cal{A}}_0(x)$ and CCFM evolution denoted by
$\bar{\cal A}(x,{\mathbf k}_{T}^2,\mu^2)$.
%\begin{eqnarray}
%x{\cal A}(x,{\mathbf k}_{T}^2,\mu^2)~=~\int dz {\cal A}_0(z)%%{x\over z} {\bar{\cal A}}({x\over z},{\mathbf k}_{T}^2,
%\mu^2). \nonumber
%\end{eqnarray}
%where
%\begin{eqnarray}
%x{\cal{A}}_0(x)~=~Nx^{p_0}(1-x)^{p_1}\exp(-{\mathbf k}_{T}^2/%k^2_0).\nonumber
%\end{eqnarray}
 The parameters of ${\cal{A}}_0(x)$ were determined in the fit to $F_2$ data.

 It is known that the hard partonic subprocess of gluon-gluon fusion $g^∗ g^∗ \to Q \bar Q$ amplitude is described by three Feynman's diagrams.
  
%\section{Numerical results}
 In the numerical calculations in the case CCFM UGD we have used two different sets, namely $A0$ and $B0$. The difference between these sets is connected with the different values of
soft cut and width of the intrinsic ${\mathbf k}_{T}$ distribution.
%A reasonable description of the $F_2$\ data
%can be achieved by both these sets.

For KMR we have used as the
standard GRV 94~(LO)~\cite{GRV} as MSTW~\cite{MSTW} (in LZ calculations) and MRST 99~\cite{MRST} (in CASCADE) sets.
The UGD dependes on
the renormalization and factorization scales  $\mu_R$ and 
$\mu_F$.
 We set 
$\mu_R^2 = m_Q^2 + ({\mathbf p}_{1T}^2 + {\mathbf p}_{2T}^2)/2$,
$\mu_F^2 = \hat s + {\mathbf Q}_T^2$, where ${\mathbf Q}_T$ is the
transverse momentum of the initial off-shell gluon pair,
$m_c = 1.4 \pm 0.1$~GeV, $m_b = 4.75 \pm 0.25$~GeV.  We use the LO formula
for the coupling $\alpha_s(\mu_R^2)$ with $n_f = 4$ active quark flavors at $\Lambda_{\rm QCD} = 200$~MeV, such that 
$\alpha_s(M_Z^2) = 0.1232$.
\begin{figure}[!b]
 \includegraphics[height=.25\textheight]{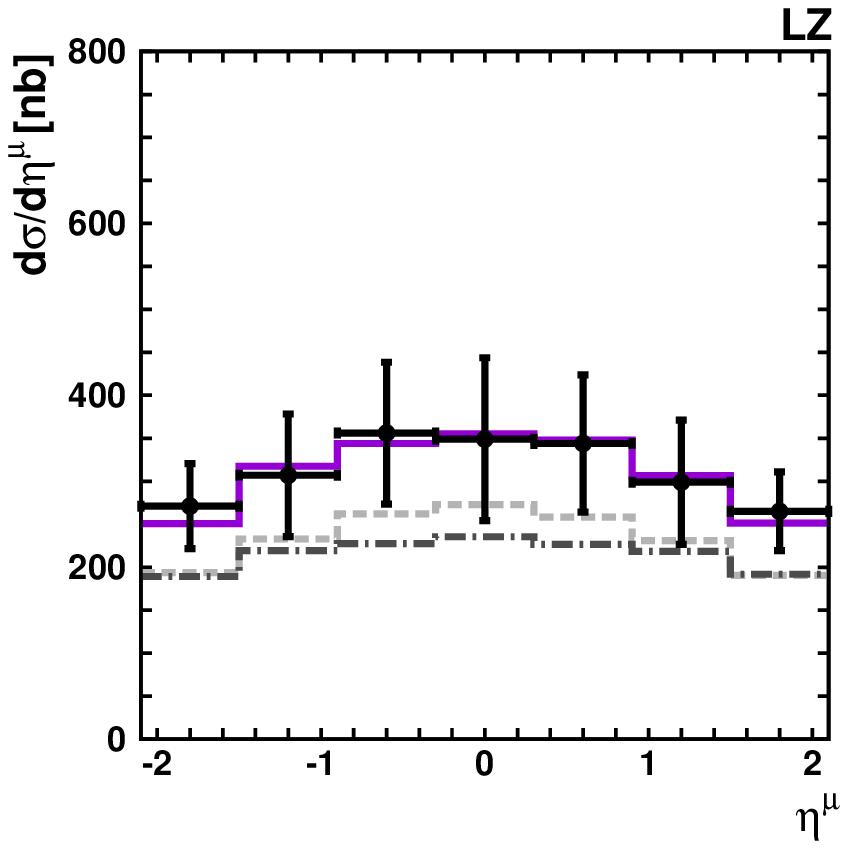}
 \includegraphics[height=.25\textheight]{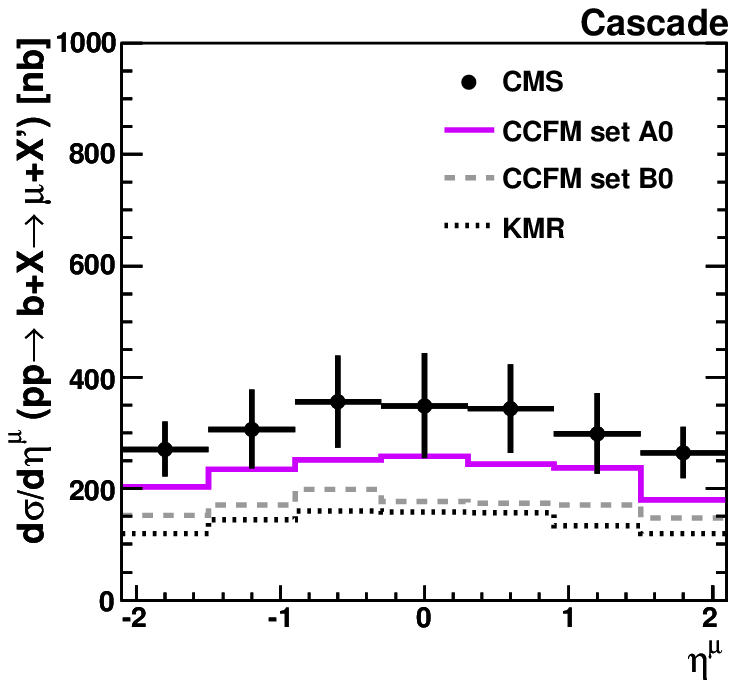}
\caption{The pseudo-rapidity distributions of muons arising from the
semileptonic decays of beauty quarks. The first column shows the LZ numerical
results while the second one depicts the CASCADE predictions.
Solid, dashed and dash-dotted, dotted histograms
correspond to the results obtained with the CCFM A0, B0
and KMR UPD.
The experimental data are from CMS.}
\label{fig1}
\end{figure}

\begin{figure}[!b]
\includegraphics[height=.23\textheight]{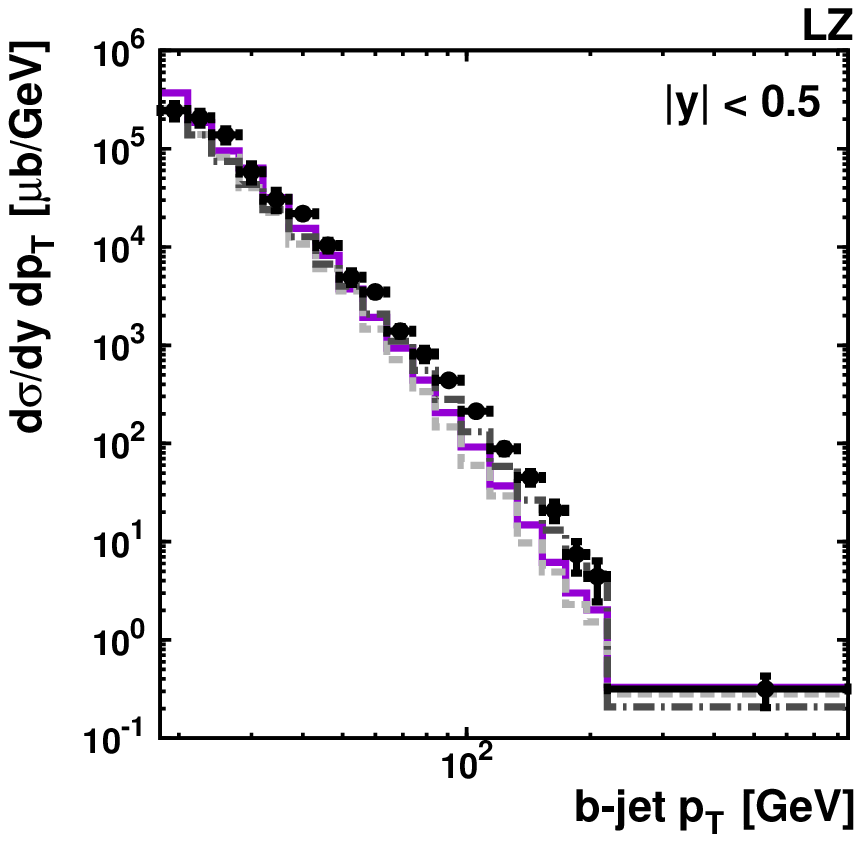}
%includegraphics[height=.27\textheight]{y-pt2-CMS.eps}
%\includegraphics[height=.27\textheight]{y-pt3-CMS.eps}
\includegraphics[height=.23\textheight]{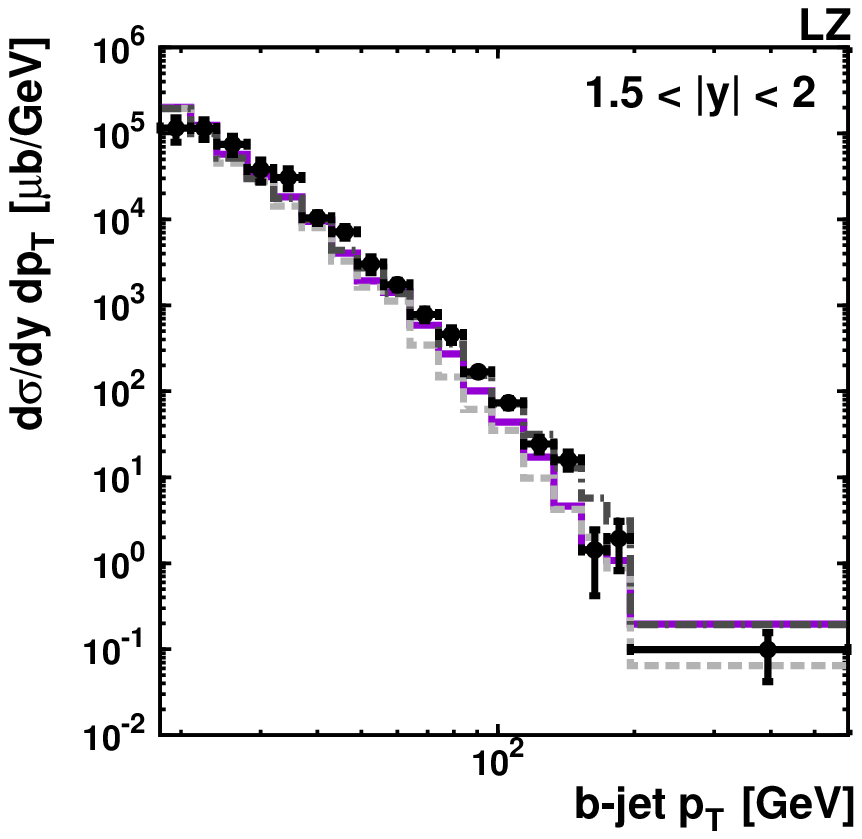}
\caption{The double differential cross sections $d\sigma/dydp_T$ of inclusive
$b$-jet production as a function of $p_T$ in different $y$ regions calculated at $\sqrt s = 7$~TeV (LZ predictions).
Notation of all histograms is the same as in Fig.~1.
The experimental data are from CMS.}
\label{fig2}
\end{figure}

\begin{figure}[!b]
\includegraphics[height=.23\textheight]{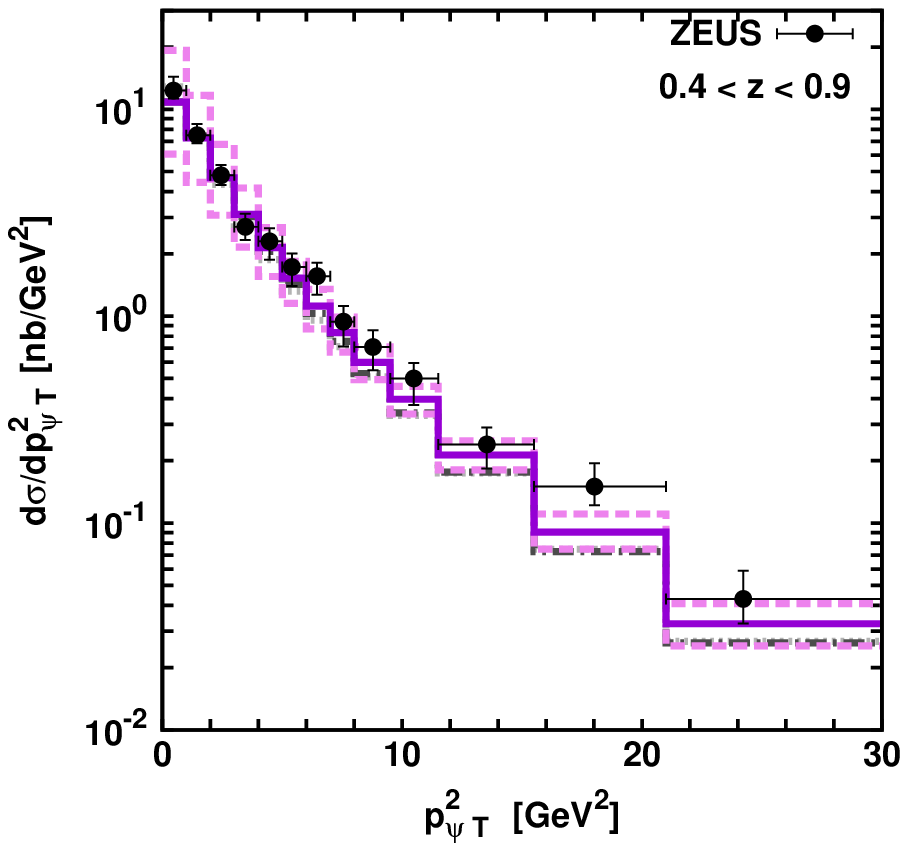}
\includegraphics[height=.23\textheight]{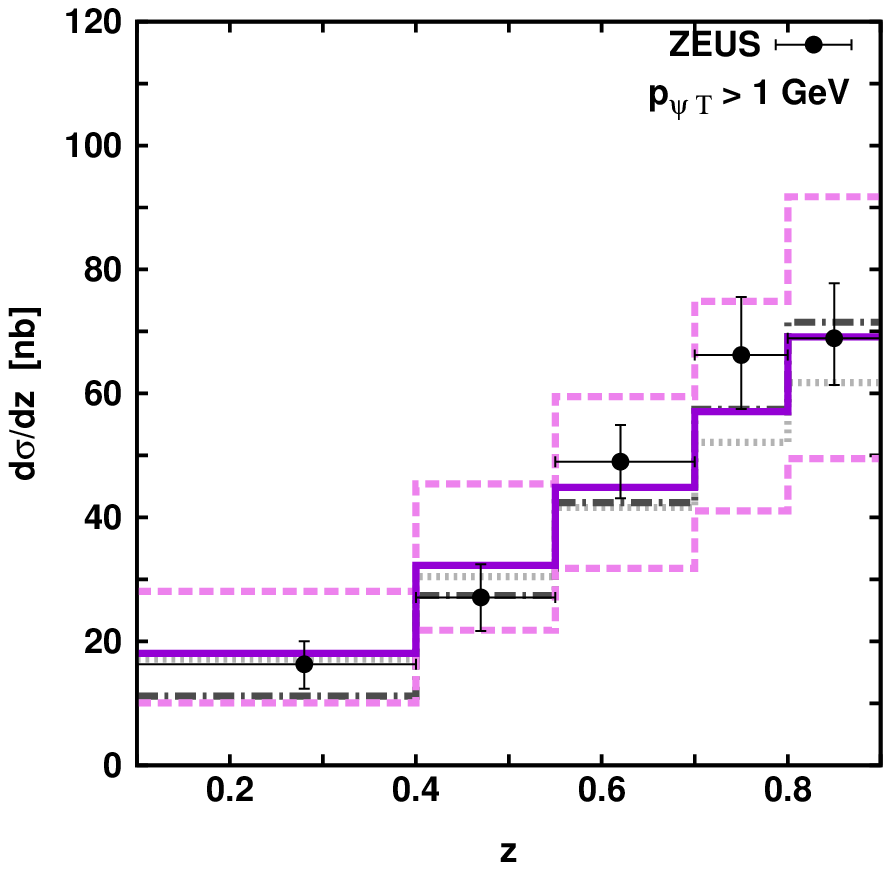}
\caption{Differential cross sections $J/\psi$ mesons
at HERA.
The solid, dashed and dash-dotted histograms correspond to
 the results
obtained using the CCFM A0, BO and KMR gluon
 densities. The upper and lower dashed histograms represent 
the scale variations.}
\label{fig3} 
\end{figure}
%\subsection{Beauty quark production in $pp$-collisions}
\begin{figure}[!b]
%\hspace*{1.2cm}
\includegraphics[height=.30\textheight]
{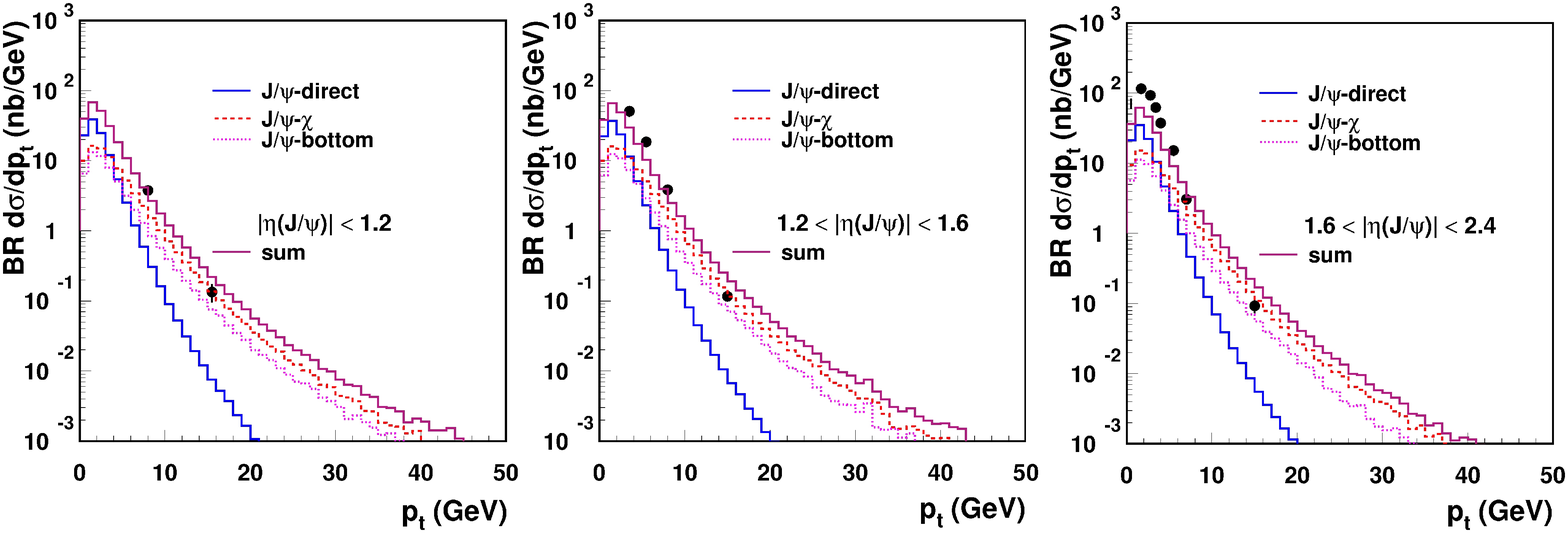}
\caption{Differential cross sections $J/\psi$ mesons
at LHC in CASCADE. The experimental data are from CMS.}
\label{figure4}
\end{figure}

We begin the discussion by presenting our results for the
muons originating from the semileptonic decays of 
$b$-quarks (Fig. 1). 
To produce muons from $b$-quarks, we first convert
$b$-quarks into $B$-mesons
using the Peterson fragmentation function with default value
$\epsilon_b = 0.006$~\cite{Peter}
and then simulate their semileptonic decay according to the standard electroweak theory taking into account the decays
 $b \to \mu $ as well as the cascade decay $b\to c\to \mu$. In CASCADE calculations also Peterson fragmentation function is used but with full PYTHIA fragmentation. In Fig. 2 we show
our discription of the $b$-jet distributions at LHC. 

%\subsection{Quarkonium production}
In the case quarkonium production we used   
Color-Singlet (CS) photon-gluon or gluon-gluon fusion in the framework of the $k_T$-factorization approach. In Fig. 3 we
show our results for $J/\psi$ production at HERA and
in Fig. 4 we present  comparison of our results with LHC data on the $J/\psi$ production in framework of the MC generator CASCADE.

In summary
 we have analysed the first data on the beauty and 
$J/\psi$ production in $pp$ collisions at LHC taken by the CMS collaboration.
Our study is based on a semi-analytical parton level
calculations and a full hadron level  MC generator CASCADE.
%\item The predictions with the default set of parameters
%tend to slightly underestimate the \Blue{$\mu$} data at
%central rapidities.
The overall description of the data is reasonable.
In most of the distributons it is similar to MC@NLO
except in some particular distributions where the 
$k_T$-factorization approach does describe the data better,
like in $b$-jet.$J/\psi$ production in the $k_T$-factorization approach with
CS model comes much closer to the data than the collinear calculations.
The reason is the off-shell ME, which includes even higher
order contributions than the NLO collinear calculations. 

The authors were supported by MSU--DESY project on MC implementation for HERA--LHC, RF FASI grant NS-4142.2010.2 and RF FASI state contract 02.740.11.0244. A.L.
 was supported in part by the grant of President of Russian Federation (MK-3977.2011.2) and the HRJRG fund. N.Z. is very grateful to the Organization Committee, in particular R. Ent, for the financial support.

\bibliographystyle{aipproc}   % if natbib is available
%\bibliographystyle{aipprocl} % if natbib is missing

%%%%%%%%%%%%%%%%%%%%%%%%%%%%%%%%%%%%%%%%%%%
%% You probably want to use your own bibtex database here
%%%%%%%%%%%%%%%%%%%%%%%%%%%%%%%%%%%%%%%%%%%

\end{document}